\documentclass[twocolumn,showpacs,
amsmath,amssymb,pre]{revtex4}

\usepackage{graphicx}

\begin{document}

\title{Mechanisms in the size segregation of a binary granular mixture}
\author{Matthias Schr\"oter}
\email{schroeter@chaos.utexas.edu}
\author{Stephan Ulrich}
\author{Jennifer Kreft}
\author{Jack B. Swift}
\author{Harry L. Swinney}
\affiliation{Center for Nonlinear Dynamics and Department of Physics, The University of Texas at Austin, Austin,
         Texas 78712, USA}
\date{\today}

\begin{abstract}
A granular mixture  of particles of two sizes that is shaken
vertically will in most cases segregate. If the larger particles
accumulate at the top of the sample, this is called the Brazil-nut
effect (BNE); if they accumulate at the bottom, the reverse
Brazil-nut effect (RBNE). While this process is of great
industrial importance in the handling of bulk solids, it is not
well understood. In recent years ten different mechanisms have
been suggested to explain when each type of segregation is
observed. However, the dependence of the mechanisms on driving
conditions and material parameters and hence their relative
importance is largely unknown. In this paper we present
experiments and simulations where both types of particles are made
from the same material and shaken under low air pressure, which
reduces the number of mechanisms to be considered to seven. We
observe both BNE and RBNE by varying systematically the driving
frequency and amplitude, diameter ratio, ratio of total volume of
small to large particles, and overall sample volume. All our
results can be explained by a combination of three mechanisms: a
geometrical mechanism called void filling, transport of particles
in sidewall-driven convection rolls, and thermal diffusion, a
mechanism predicted by kinetic theory.
\end{abstract}


\maketitle

\section{Introduction}
 A variety of mechanisms have been proposed to describe the
separation of particles of two sizes in a mixture of vertically
shaken particles.  Section II discusses proposed mechanisms,
including void filling, static compressive force, convection,
condensation, thermal diffusion, non-equipartition of energy,
interstitial gas forcing, friction, and buoyancy. Our aim is to
determine which are the important segregation mechanisms for
diverse conditions. 

We study a mixture of two sizes of brass spheres, using image
analysis to  count the number of large particles visible at the
top and bottom surfaces of the sample after shaking. These results
are then compared with molecular dynamics simulations, which yield
quantities unaccessible in experiment. Comparing our results for a
range of control parameters with various proposed segregation
mechanisms reveals which mechanisms are important in our system.
The control parameters studied are: frequency $f$ and amplitude
$A$ of the sinusoidal shaking, the diameter ratio of the spheres
$d_{\rm L}/d_{\rm S}$ (the subscripts L and S refer to large and
small particles), the total layer depth $h_{\rm total}$, and
$V_{\rm L}/V_{\rm S}$, the ratio of the total volume of large to
small particles in the container.  The last parameter has not
been examined in previous experiments, which were either performed
with about equal numbers of particles of each size
\cite{olsen:64,faiman:65,hsiau:97,brone:97,burtally:02,breu:03}, 
or in the case
called the intruder limit, where there is one large sphere in a
bed of smaller particles
\cite{williams:63,duran:93,duran:94,knight:93,knight:96,cooke:96,vanel:97,shinbrot:98,liffman:01,fernando:03,nahmad:03,huerta:04,yan:03,moebius:01,moebius:04,moebius:05,huerta:05}.

This paper is organized as follows:
different proposed segregation mechanisms
are reviewed in Sec.~\ref{sec:mechanism}. Section \ref{sec:theory} discusses
the interplay of these mechanisms in two recent models \cite{hong:01,trujillo:03}.
The experimental setup and the details of our simulation are described in
Sec.~\ref{sec:methods}.
Section \ref{sec:results_1} discusses how the
experimental and simulation results depend on the driving parameters, and
Sec.~\ref{sec:results_2} discusses the dependence on  $d_{\rm L}/d_{\rm S}$, $V_{\rm L}/V_{\rm S}$
and  $h_{\rm total}$.
We conclude in Sec.~\ref{sec:conclusions}.

\section{Segregation Mechanisms}
\label{sec:mechanism} To reduce the number of segregation
mechanisms to be considered we use large and small particles of
the same material (brass) and therefore the same density and
mechanical properties. This excludes segregation effects due to
Archimedean buoyancy
\cite{shishodia:01,trujillo:03,huerta:04,bose:05,ciamarra:05,huerta:05}
or due to different frictional properties
\cite{srebro:03,ciamarra:05}. As the interstitial fluid was also
found to have a strong effect on  segregation
\cite{burtally:02,biswas:03,yan:03,moebius:01,moebius:04,moebius:05},
we reduced the pressure in our experiment to less than 8 Pa, well
below the 130 Pa where the effect of air on the segregation
vanishes \cite{moebius:04}. This step also assures comparability
with our molecular dynamics simulations where the only
interactions between particles are collisions. The remaining seven
mechanisms are reviewed below, and their predicted
dependence on $d_{\rm L} / d_{\rm S}$, $V_{\rm L} / V_{\rm S}$ and
$h_{\rm total}$ is summarized in Table \ref{tab:mechanism}. For a
recent review see also \cite{kudrolli:04}.

\subsection{Geometrical mechanism}
\subsubsection{Void filling}
\label{sec:void_filling}
\label{sec:dynamic_tensile}
Void filling is a local geometric effect that was first clearly identified in Monte Carlo simulations \cite{rosato:87}.
During the expansion phase of the shaking cycle all particles move upwards,
but during the compaction phase a large particle has a smaller
probability of finding a suitable void in the layer below it than a small particle.
This leads to the large particles accumulating at the top of the sample.

As pointed out in \cite{kudrolli:04}, void filling works only  for weak excitation;
for strong driving the voids become larger and the probability that they accommodate a large
particle becomes similar to that of a small particle.
The void filling effect becomes stronger with increasing
diameter ratio $d_{\rm L} / d_{\rm S}$, as shown by the increase in the
rise velocity of an intruder \cite{vanel:97}.
For $d_{\rm L} / d_{\rm S} \approx 3$ the intruder seems to undergo a transition from intermittent to continuous
ascent,
which indicates a change in the microscopic details of the mechanism \cite{duran:93,duran:94}.
Comparing in a thought experiment the situation of a single large intruder in a bed of small particles
with a mixture which
contains predominantly large particle hints that increasing $V_{\rm L} / V_{\rm S}$ will decrease the strength of
void filling. In contrast, the overall layer height $h_{\rm total}$ should not influence this mechanism.

There exists no theoretical model for void filling, but
Trujillo \emph{et al.}~\cite{trujillo:03} identify in their
kinetic theory based analysis of the gravitational force
a  ``dynamic tensile'' force; for an intruder this upward force
increases with $(d_{\rm L} / d_{\rm S})^3$.
This term  vanishes in the limit of a highly expanded mixture.
They compare the dynamic tensile force with the void filling mechanism.

\subsubsection{Static compressive force}
\label{sec:static_comp} Trujillo \emph{et al.}~\cite{trujillo:03}
find  that in the presence of gravity, any size disparity will
give rise to another term, which they name
static compressive force. The static compressive force will always
lead to the RBNE; in the intruder limit it is proportional to
$(d_{\rm L} / d_{\rm S})^3 - 1$.

\subsection{Convection}
\label{sec:convection} Friction at the container walls can induce
convection rolls in vertically shaken mixtures: during the upward
acceleration the mixture gets compacted and shear forces induced
by the side walls extend through the whole sample. During the
subsequent downward motion the mixture is more expanded and
consequentially those particles adjacent to the walls experience
stronger shear forces than those in the center of the container.
Both phases combined give rise to a convection roll going
downwards at the sidewalls and upwards in the center.

The relevant geometrical control parameter seems to be the ratio
of the size of the downstream layer to the diameter of the large
particles. If the large particles are too large to be entrained by
the downstream, they become trapped at the top surface and
consequently convection leads to BNE
\cite{knight:93,knight:96,cooke:96}. However, if the size of the
downward moving regions is big enough to accommodate large
particles, these particles perform convective cycles
\citep{knight:93,poeschel:95,shinbrot:98,rosato:02,nahmad:03},
which can lead to mixing \cite{brone:97}.

Magnetic resonance imaging indicates that the size of the
downward moving layer is independent of total layer depth and shaking acceleration \cite{knight:96}.
However, increasing $h_{\rm total}$ might reduce the efficiency of both mechanisms as the
strength of the convection roll
vanishes at the bottom \citep{poeschel:95}.
Increasing the ratio of large to small particles  $V_{\rm L} / V_{\rm S}$ in a gedanken-experiment \cite{kudrolli:04}
increases the size of the downstream layer and therefore suppresses the BNE and fosters mixing.

\subsection{Kinetic energy  driven mechanisms}
\label{sec:kin_energy_mechanism} For driven granular media one can
define a kinetic granular temperature $T$ which corresponds to the
average kinetic energy of the random motion of the particles
\cite{brilliantov:04},
\begin{equation}
T=\frac{m}{3} \langle (v-\langle v\rangle)^2 \rangle,
\label{eq:t}
\end{equation}
where $m$ is the particle mass and $v$ their velocity. Numerical
studies  \cite{soto:99,bougie:02,moon:04} and experiments
\cite{wildman:02,wildman:03,huan:04} have shown  that $T$ in a
vertically shaken granular sample depends not only on the position
in the sample and the phase of the shaking cycle, but also on the
shaking acceleration and the number of particles in the container.

In binary granular mixtures the equipartition of energy breaks down
\cite{losert:99,wildman:02,wildman:03,feitosa:02}; in general
the heavier particles have a higher granular temperature than the lighter ones.
Non-equipartition increases with decreasing restitution coefficient and increasing mass
difference \cite{galvin:05}.

Two different types of segregation mechanisms driven by the granular temperature of the sample have been suggested,
based either on a phenomenological model of solid-liquid phase transitions \citep{hong:01}
or on analyzing momentum balance equations derived from kinetic theory
\cite{hsiau:96,jenkins:02,garzo:02,trujillo:03,galvin:05,brey:05,garzo:06}.

\subsubsection{Condensation}
\label{sec:condensation}
A mechanism suggested by Hong {\it et al.}~\citep{hong:01} is based on the
observation that as the  shaking acceleration is increased,  a solid-like sample of {\it monodisperse} particles
will become vibro-fluidized. Assuming
that the granular temperature is spatially
homogeneous throughout the sample, they postulate the existence of a
critical temperature $T^{\rm c}$ above which all particles are fluidized.
For $T$ smaller than $T^{\rm c}$ an increasing fraction of the particles
should condense at the bottom of the container.
By equating the average kinetic energy of the particles
with the potential energy of a particle in the top layer,
they find  $T^{\rm c}$ to be proportional to $\rho d^3 h$,
where $\rho$ is the material density and  $h$ is the height of the particular layer.

Hong {\it et al.}~\citep{hong:01} argued that in a {\it binary}
sample the shaking parameters can be adjusted such that $T$ is
between the critical temperatures of the two species. Then one
kind stays fluidized while the other condenses at the bottom.
Assuming equipartition and that  the values of  $T^{\rm c}$ are
not influenced by the presence of the other species, they identify
the ratio of the two $T^{\rm c}$'s as the  control parameter
$\epsilon_{\rm cond}$,
\begin{equation}
        \epsilon_{\rm cond} = \frac{T^{\rm c}_{\rm L}}{T^{\rm c}_{\rm S}}
                            = \left ( \frac{d_{\rm L}}{d_{\rm S}} \right )^3
                               \frac{\rho_{\rm L}}{\rho_{\rm S}}   \frac{h_{\rm L}}{h_{\rm
                               S}},
\label{eq:epsilon_cond}
\end{equation}
where the height ratio $h_{\rm L}/h_{\rm S} $  of the particles filled in separately
is equal to  $V_{\rm L}/V_{\rm S}$ in a container of constant width.

For $\epsilon_{\rm cond} < 1$, which corresponds to  $T^{\rm c}_{\rm L} < T^{\rm c}_{\rm S}$,
the small particles are  still condensed and sink to the bottom while the large particles are completely fluidized;
the system shows BNE.
For  $\epsilon_{\rm cond} > 1$ the large particles will condense at the bottom and the RBNE takes place.
However, these effects will  only occur in an intermediate driving regime where
$T$ is in between the two critical temperatures;
otherwise there will be no segregation as a result of condensation.

\subsubsection{Mechanisms predicted by kinetic theory}
\label{sec:thermal_mech}
Kinetic theory of binary granular mixtures  is an active area of research
with competing predictions.
As kinetic theory is limited to a regime where
binary collisions prevail, these mechanisms are only
applicable for dilute mixtures and hence large shaking amplitudes.

{\bf Diffusion in a granular temperature gradient}. It was first
shown by Hsiau and Hunt \cite{hsiau:96} that in a gradient of the
granular temperature the  heavier particles move to places with
lower $T$ and thereby expel the lighter particles. Strong vertical
temperature gradients in vertically shaken samples have been found
in simulation \cite{soto:99,bougie:02,moon:04} and experiment
\cite{feitosa:02,huan:04}. Due to the complicated spatio-temporal
development of the $T$ field, it is not possible to decide a
priori what type of segregation this effect will lead.

{\bf Segregation due to non-equipartition}. The derivation in
\cite{hsiau:96} assumes equipartition. Recent simulations by
Galvin {\it et al.}~\cite{galvin:05} have shown that
non-equipartition  leads to additional driving forces that  are
comparable in strength to the forces in the calculations assuming
equipartition.

An explicit solution for the thermal diffusion coefficients in the case of a
single intruder and non-equipartition was given by Brey {\it et al.}~\cite{brey:05},
who found the ratio of mean square velocities,
\begin{equation}
        \phi = \frac {m_{\rm S} T_{\rm L}} {m_{\rm L} T_{\rm S}} ,
\label{eq:brey}
\end{equation}
to be the control parameter.
For particles with same density, they predict $\phi < 1$, which leads to RBNE;
$\phi$ decreases with increasing diameter ratio.

Trujillo \emph{et al.}~\cite{trujillo:03} identify in their
hydrodynamic model a so-called pseudo-thermal buoyancy force,
which is proportional to the ratio of the granular temperatures of
the two species. They calculate this temperature ratio using the
kinetic theory results  of Barrat and Trizac \cite{barrat:02},
where $T_{\rm L}$ is always larger than $T_{\rm S}$ so that the
pseudo-thermal buoyancy leads to BNE. Barrat and Trizac also find
that non-equipartition increases strongly with $d_{\rm L}/d_{\rm
S}$ and decreases weakly with $V_{\rm L}/V_{\rm S}$.

\begin{table}
\caption{Effect of  an increase of the diameter ratio, volume ratio and total layer depth
        on the segregation mechanism discussed in Sec.\ref{sec:mechanism}.
    The symbols +, 0, and  - correspond to an increase, constance or decrease of the effect in the second column.}
\label{tab:mechanism}
\begin{tabular}{c||c|c|c|c}
mechanism  & effect & incr. $\frac{d_{\rm L}}{d_{\rm S}}$ & incr. $\frac{V_{\rm L}}{V_{\rm S}}$ & incr. $h_{\rm total}$\\
\hline
\hline
void filling  & BNE      & +  & - &  0\\
\hline
static compressive  & RBNE & +  &  ?  & ? \\
force & & & & \\
\hline
convection    & BNE      & +  &  -  &   -       \\
              & mixing   & -  &  +  &   -      \\
\hline
condensation  & BNE      & -  & -  & 0    \\
              & RBNE     & +  & +  & 0  \\
\hline
thermal diffusion & ?     &?   & ?  &  ?   \\
\hline
non-equipartition   & RBNE & + & ? &  0  \\
Brey {\it et al.} & & & & \\
\hline
non-equipartition   & BNE & + & - &  0  \\
Trujillo {\it et al.} & & & & \\
\end{tabular}
\end{table}

\section{Combination of mechanisms}
\label{sec:theory}
To predict the state of segregation in an experiment
the different mechanisms in Sec.~\ref{sec:mechanism} have to be
compared in their relative strength for a given set of control parameters.
Here we focus on the phenomenological model from
Hong {\it et al.}~\cite{hong:01}
and the approach based on kinetic  theory  from
Trujillo \emph{et al.}~\cite{trujillo:03}.

We could not compare our results to predictions of hard sphere
lattice models
\citep{nicodemi:02,coniglio:04,tarzia:04,tarzia:05}, which are
based on  a statistical mechanics approach to granular media
\cite{edwards:89}. Their main control parameter is a
configurational temperature, which we were unable to determine in
our experiments and simulations.

\subsection{The model of Hong, Quinn and Luding}
\label{sec:hql}
While Hong {\it et al.} do not deny the presence of other segregation mechanisms
in specific situations, they propose that a theory based on void filling (percolation in their nomenclature,
discussed in Sec.~\ref{sec:void_filling})
and the condensation mechanism (Sec.~\ref{sec:condensation}) will provide the general frame work
to predict if the BNE or the RBNE is observed.
Interpreting the  results from Rosato et al.~\cite{rosato:87}, they assume void filling to be controlled by
$\epsilon_{\rm vf}$
\begin{equation}
        \epsilon_{\rm vf}      = \left ( \frac{d_{\rm L}}{d_{\rm S}} \right )^3
\label{eq:epsilon_vf}
\end{equation}

Equating $\epsilon_{\rm vf}$ and $\epsilon_{\rm cond}$ (Eq.~\ref{eq:epsilon_cond})  results in
an equation for the phase boundary between the two effects which is
\begin{equation}
       \frac{h_{\rm L}}{h_{\rm S}} =  \frac{\rho_{\rm S}}{\rho_{\rm L}}
\label{eq:HQL_1}
\end{equation}
with the RBNE if the left hand side is larger than the right hand side and the BNE for the opposite case.
Hong {\it et al.} study however the more special case
where the layer height ratio is chosen to be equal to the diameter ratio
$h_{\rm L} / h_{\rm S} = d_{\rm L} / d_{\rm S}$ .  Consequently they obtain for the phase boundary
\begin{equation}
        \frac{d_{\rm L}}{d_{\rm S}}  =  \frac{\rho_{\rm S}}{\rho_{\rm L}},
\label{eq:HQL_2}
\end{equation}
which is in agreement with their molecular dynamics simulations \cite{hong:01}
and a variational approach minimizing a binary mixture free energy functional \cite{both:02}.

An experimental test of Eq.~\ref{eq:HQL_1} or \ref{eq:HQL_2} is complicated by the fact that
the condensation mechanism can only be observed for an intermediate granular temperature,
which is difficult to achieve in experiment.
Consequently Breu {\em et al.} \cite{breu:03} did a first study of Eq.~\ref{eq:HQL_2}
by manually preparing binary mixtures in the layer sequence predicted to be unstable for this
diameter and density ratio.
Then they searched for a combination of the control parameters
$f$ and $A$ where the configuration changed to that predicted.
This search succeeded for 82\% of the 178 different particle combinations tested.
However, they also observed that for
the same mixture both BNE and RBNE can occur, depending on the driving parameters.
Other experimental \cite{canul:02,moebius:05} and numerical \cite{ciamarra:05}
studies could not verify the existence of the condensation mechanism.

\subsection{The model of Trujillo, Alam and Herrmann}
\label{sec:trujillo}
All segregation mechanisms discussed in this paper originate from the presence
of a gravitational field parallel to the direction of shaking; in the framework
of kinetic theory the influence of this field was first studied by Jenkins and Yoon \cite{jenkins:02}.
Starting from the same momentum balance equations, Trujillo \emph{et al.}~\cite{trujillo:03}
decompose the net gravitational force into four physical mechanisms.
Two of them are geometric in nature: the static compressive force (discussed
in Sec.~\ref{sec:static_comp}) and the dynamic tensile force,
which they compare with the void filling mechanism
(see Sec.~\ref{sec:dynamic_tensile}). The third term stems from the
non-equipartition of energy and is proportional to $T_{\rm L} /
T_{\rm S} $; it is called pseudo-thermal buoyancy
(Sec.~\ref{sec:thermal_mech}). Finally, there is the classical
Archimedean buoyancy, which is not relevant for particles of same material densities.
The model of  Trujillo \emph{et al.} does  not account for
convection or spatial inhomogeneities of the kinetic temperature in the sample.

For  the two-dimensional case they find that for small $d_{\rm L} / d_{\rm S}$ the
static compressive force, which leads to RBNE, is stronger than
the other two terms, which favor BNE. However, the pseudo-thermal buoyancy grows faster with
increasing $d_{\rm L} / d_{\rm S}$, so there exists a critical diameter ratio above which the
system exhibits  BNE.
They also predict that increasing the fraction of large particles will
strongly increase the range of diameter ratios where the RBNE is observed.
The same effect should apply for lowering the packing fractions by means of stronger shaking.

\section{Methods}
\label{sec:methods}

\subsection{Experimental setup}
\label{sec:experiment}

\begin{figure}[t]
  \begin{center}
    \includegraphics[width=8cm]{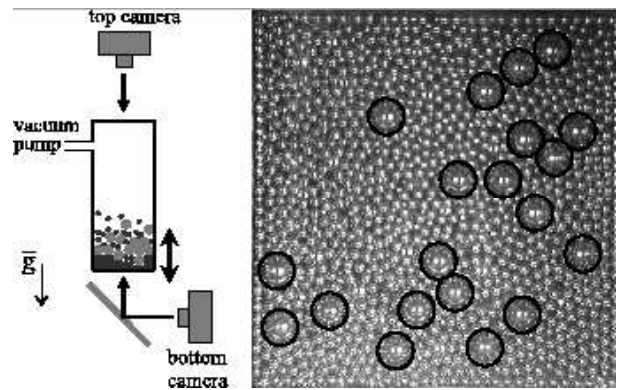}
    \caption{Left: Experimental Setup. The particles are shaken vertically in an
      evacuated square glass tube.
        Pictures are  taken from the top and  the bottom (using a mirror) of the
        granular sample.
      Right: An image of the bottom plate. All large particles detected by the image processing
        are marked with black circles. }
    \label{setup}
  \end{center}
\end{figure}

The experimental Setup is shown in Fig.~\ref{setup}. The particles are shaken in
a 25\;mm x 25\;mm square glass tube (Pyrex$^\circledR$ Borosilicate) with a height
of 360\;mm. The bottom plate is made out of sapphire, the top cover
of acrylic glass.
The cell is mounted on an electromechanical shaker (Vibration Test Systems vg100c)
and shaken sinusoidally with an amplitude $A$ and frequency $f$.
The dimensionless shaking acceleration is $\Gamma = (2 \pi f)^2 A / g$,
where $g$ is the acceleration due to gravity.
$\Gamma$ is controlled by a feedback loop to better than 0.4\%.

All particles were made of brass (alloy 260) with a density $\rho$ of 8.4\,${\rm g/cm^3}$.
The diameter  $d_{\rm L}$ of the large particles was 2.38\,mm;
the small particle diameters  $d_{\rm s}$ were 0.79\,mm, 1.19\,mm, and 1.59\,mm,
corresponding to diameter ratios of 3, 2 and 1.5.

All experiments were performed with the same protocol:
after filling with particles, the container was sealed and evacuated  to a pressure below 8 Pa.
Then a sequence of measurements was performed, each corresponding to a specific set of driving parameters
$f$ and $\Gamma$. Each individual measurement started with the generation
of an approximately mixed state by shaking 30\,s at $f=15$ Hz and $\Gamma = 4.5$.
Then the particles were shaken with the specific $f$ and $\Gamma$ until a steady state was reached.
After the shaker was stopped the particles came immediately to rest and
images were taken from the top and bottom surface of the sample
with a resolution of 40 pixels per  $d_{\rm L}$.
Three more sets of images were  taken, each after shaking for another 20\,s.
The average number of particles counted in the four sets of images was considered one independent measurement.

To identify the position and type of all particles we use two images  taken with different illumination.
A first image, taken with a single point light source, marks the center of
each particle with a light spot. A second image, taken with diffuse illumination, is then used to distinguish
between small and large particles. The algorithm  makes this decision automatically by
comparing the  correlation between every found particle and template images of
small and large particles.
The right side of Fig.~\ref{setup} shows a result of this image processing. The error rate of the algorithm,
as determined by comparing the number of large particles to the results of counting them by eye, is smaller than 6\%.

\begin{figure}[t]
  \begin{center}
    \includegraphics[width=8.5cm]{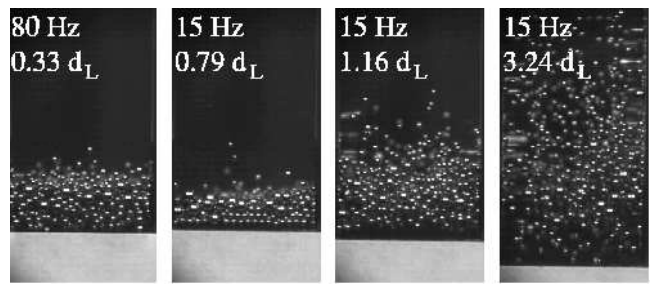}
    \caption{Side view of the shaken sample while it is most expanded. The labels give the shaking
    frequency and amplitude, corresponding to $\Gamma$ = 20, 1.7, 2.5, and 7, respectively.
     The container is filled with 128 large and approx.~1040 small brass spheres
    (corresponding to equal volumes); their diameter ratio is  2.
    The elongated appearance of the particles in the horizontal
    direction is a consequence of the  illumination. The height of the images is
  	one-eighth of the overall tube length; this height is large enough so
  that the spheres do not collide with the top of the container.
        }
    \label{fig:high_speed}
  \end{center}
\end{figure}

To quantify the amount of convection we took additional images of a sidewall of the sample during the shaking process.
Using a Phantom high speed camera we resolve the particle dynamics with a resolution of 24 pixels per  $d_{\rm L}$
at 1077 frames per second. The high frame rate allowed identification of the motion of individual particles.
A small depth of field helped to distinguish particles near the sidewalls from particles in the bulk.
Figure \ref{fig:high_speed} gives four examples.

\begin{figure}[h]
  \begin{center}
    \includegraphics[angle=-90,width=8.5cm]{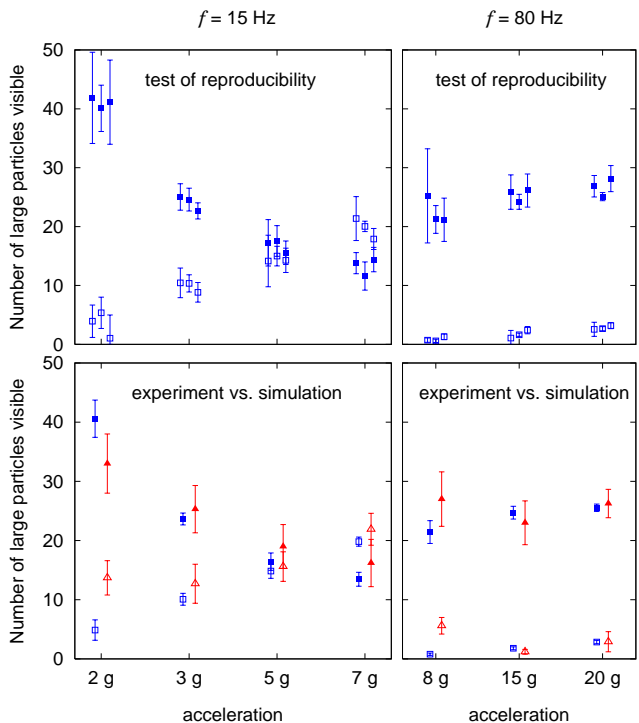}
    \caption{
    (Color online) Top row: Test of the reproducibility in three experimental
    runs, each starting with washed particles as described in Sec.~\ref{sec:experiment}.
    Open symbols correspond to the average number of large particles detected  at the bottom;
    closed symbols to the number on the top. Each set of points represents experiments with identical
    shaking acceleration. Bottom row: Comparison of average of the experimental results in the top row (squares)
    and the molecular dynamics simulations (triangles). The sample contains 64 large and 1540 small brass spheres with a
        diameter ratio of 2.
        }
    \label{fig:repro}
  \end{center}
\end{figure}

In order to get reproducible results the brass particles had to be
treated in two ways. First, all new particles were aged by shaking
them for 24 hours at 20 Hz and $\Gamma$ = 5 to exclude effects due
to work hardening. Second, after about every 8 hours of shaking
the particles were cleaned for 10 minutes in an ultrasonic bath
with a concentrated solution of Alconox powdered cleaner. This
removed  the  fine metal dust which had accumulated in the sample.
The resultant reproducibility is illustrated in the top row of
Fig.~\ref{fig:repro}

\subsection{Molecular dynamics simulation}
\label{sec:MD}

A molecular dynamics code described in \cite{bizon:98} is used for
the simulations. In the simulation the container size, number of
particles, diameter ratio, mass ratio, and shaking parameters are
all identical to those in the experiment. Collisions are assumed
to be binary and instantaneous to permit the use of an event
driven algorithm to decrease calculation time. The velocities
before and after an event are calculated using the collision model
proposed by Walton, which does not take into account the details
of an interaction but does account for friction and rotation
\cite{walton:93}.

For both particle-particle and particle-wall collisions, three
constants each (a total of 6) must be defined in this model.
  First is the coefficient of normal restitution, $e$, which is the
  ratio of the relative normal velocity ($v_n$) after the collision to the value before the collision.
This coefficient in both cases varies with $v_n$ as described in
\cite{bizon:98}: the coefficient of restitution is the maximum of
$e$ and $1-(1-e)(v_n/\sqrt{gd})^{3/4}$. The changes in relative
surface velocity are determined by the coefficient of friction
$\mu$ and the minimum coefficient of tangential restitution
$\beta$. The ratio of the value after to before the collision is
the maximum of $-\beta$ and $\mu$ times the normal impulse. This
allows a transition from sliding to rolling contact.

\subsection{Comparison of experiment and Molecular dynamics simulation}
\label{sec:comparison} The collision parameters were
systematically varied until optimal agreement was obtained between
simulation and experiment; the results are shown in
Fig.~\ref{fig:repro} lower row. This led to a choice of
coefficient of normal restitution, $e=0.78$, and friction
coefficient, $\mu = 0.3$, for both particle-particle and
particle-wall collisions.

For  particle-particle interactions, the minimum tangential
restitution, $\beta_p$, was set to 0.5; the results were only
weakly dependent on $\beta_p$. Critical to the match was
$\beta_w$, the tangential restitution for particle-wall
interactions, which was set to 0.9. This sensitivity to $\beta_w$
hints towards the importance of sidewall-driven convection in the
system. For a detailed discussion of the significance of $\beta$
in this collision model, see \cite{luding:98}.

To mimic the experiment, $\Gamma$ was reduced to $0.5$ after
1000 cycles for simulations at 15 Hz and 40,000 cycles for 80 Hz.
In the event drive simulation the particles did not fully come to
stop as quickly as they did in the experiment. The simulation
continued until the time between collisions with the plate became
too small to be handled by the algorithm. For $\Gamma=2$, $f$ = 15
Hz and $\Gamma=8$, $f$ = 80 Hz, this collapse happened quickly as
the layer was dense and close to the plate. The results from
these simulations differ from experiment more than for higher
shaking amplitudes as some of the dynamics involved in the layer
coming to rest are not captured.

The number of particles on top and bottom were determined in
simulation from the heights of the particles. If the center a
large particle was within $0.7d_{\rm L}$ of the base, it was considered
to be in the reverse state; if more than $0.3d_{\rm L}$ above the mean
height of the small particles, in the Brazil nut state. This
measure was determined by making images of the simulation results
with the ray-tracing program Povray \cite{povray},
 which were then analyzed in the same
fashion as in the experiment. The cutoff values used minimized the
difference between the two methods.

\section{Results I: Influence of driving frequency and amplitude}
\label{sec:results_1}

In this section we will discuss the influence of the driving parameters on
the segregation in a ``typical'' sample consisting of each one monolayer (measured in units of $d_{\rm L}$)
of small and large particles with a diameter ratio of two:
$d_{\rm L} / d_{\rm S}$ = 2, $V_{\rm L}/V_{\rm S}$ = 1, and  $h_{\rm total}$ = 2 $d_{\rm L}$.
We will first describe our experimental results and then
use our molecular dynamics simulations
to understand the role of convection and granular temperature inside the sample.
In Sec.~\ref{sec:results_mechanism} we will then explain our findings
in terms of the segregation mechanism discussed in Sec.~\ref{sec:mechanism}.
Section \ref{sec:results_2} examines how these results change with the diameter ratio,
the fraction of large particles and the overall filling height.

\subsection{Relaxation into a steady state}
\label{sec:steady_state}
\begin{figure}[t]
  \begin{center}
    \includegraphics[angle=-90,width=8cm]{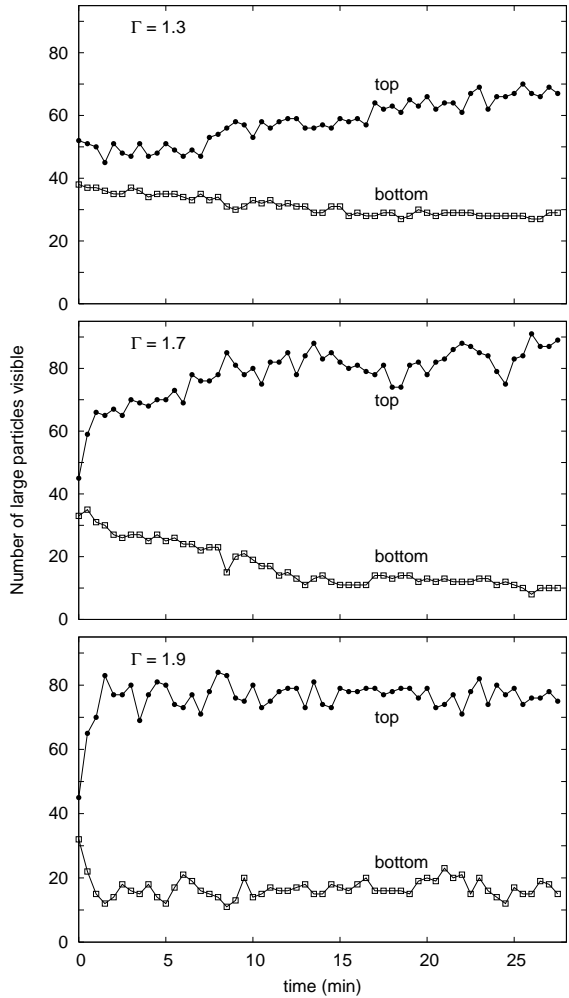}
    \caption{Relaxation into a steady state exhibiting the BNE.
    The shaking frequency in the experiment is 15 Hz.
         The container is filled with 128 large and approximately 1040 small brass spheres
        (corresponding to equal volumes); their diameter ratio is  2.
        }
    \label{fig:relax}
  \end{center}
\end{figure}
An assumption of the measurement protocol described in
Sec.~\ref{sec:experiment} is that the initial shaking period of 1
min is long enough for the system to reach a steady state. This is
demonstrated in Fig.~\ref{fig:relax} where the shaker was stopped
every 30\,s to take images and count the number of large particles
visible at the bottom and top surface. The time the sample needs
to relax to  a steady state decreases strongly with increasing
acceleration. In experiments with $\Gamma \geq 2.5$ at 15 Hz and
$\Gamma \geq 8$ at 80 Hz we reach a steady state in less than 1
minute, for lower accelerations we increased the shaking time
accordingly. We also find this steady state in agreement with
\cite{olsen:64,faiman:65} to be history independent: Starting from
samples prepared in a BNE or RBNE configuration results in the
same average number of particles.

\subsection{Dependence on the driving amplitude}
\label{sec:driving_parameters}

\begin{figure}[t]
  \begin{center}
    \includegraphics[angle=-90,width=8.5cm]{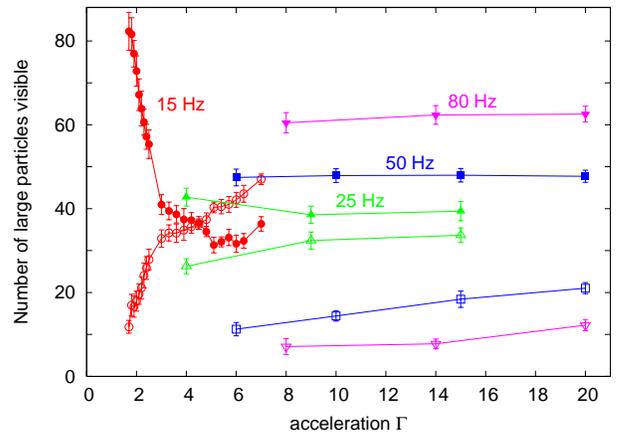}
    \caption{(Color online) Experimental segregation results as a function of
        the shaking acceleration. The different symbols correspond to different shaking frequencies;
        the open version of a symbol represents
        the average number of large particles visible at the bottom, the
        closed version the number at the top.
    These experiments exhibit the BNE except for those with $f$ = 15 Hz and $\Gamma$ $>$ 4.
        The container is filled with 128 large and approx.~1040 small brass spheres
        (corresponding to equal volumes); their diameter ratio is  2.
        All data points for $\Gamma > 2.5$ are averaged over 19  independent experiments.
        For the measurements at $1.7 <\Gamma <2.5$ we used relaxation curves of the type shown in Fig.~\ref{fig:relax},
        where the initial shaking time necessary to reach a steady state (up to 15 minutes) was discarded.
        }
    \label{fig:driving}
  \end{center}
\end{figure}

Figure \ref{fig:driving} demonstrates that the segregation behavior is relatively complicated when plotted
as a function of the shaking acceleration.
For  $f$ = 15 Hz there is a crossover from a strong BNE to a weak RBNE at $\Gamma \approx 4.5$.
In contrast for higher values of $f$ we find the steady state always to be BNE and largely independent of $\Gamma$.

To characterize the segregation behavior we introduce the probability
that an observed large  particle is at the top surface, $p = n_{\rm t}/ ( n_{\rm t} +  n_{\rm b})$,
where $ n_{\rm t}$ and $ n_{\rm b}$ correspond to the average number of large particles
detected at the top and bottom surfaces.
Consequently $p=1$ identifies a complete BNE,  $p=0$ a RBNE state.

\begin{figure}[t]
  \begin{center}
    \includegraphics[angle=-90,width=8cm]{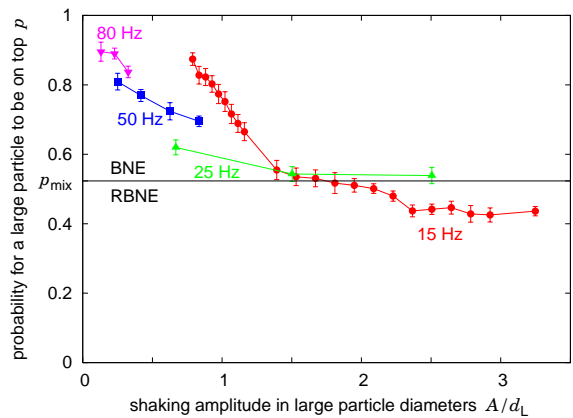}
    \caption{(Color online) Experimental segregation results as a function of the shaking amplitude.
        The data presented are the same as in Fig.~\ref{fig:driving}.
        ($p_{\rm mix}$ = 0.523 \cite{endnote})
        }
    \label{fig:amplitude}
  \end{center}
\end{figure}

The results for the probability $p$ as a function of the driving amplitude $A$ are
 shown in Fig.~\ref{fig:amplitude}:
with increasing shaking amplitude the system passes from a strong BNE to a weak RBNE.
Qualitatively similar results have been found in the experiments of Breu {\it et al.}~\cite{breu:03}
and the soft-core molecular dynamics simulations of Pica Ciamarra {\it et al.}~\cite{ciamarra:05}.

Plots of $p$ as a function of $\Gamma$ (compare Fig.~\ref{fig:driving}),
the average plate velocity or the particle velocity at lift off
do not collapse the data as well as the dependence on $A$ (Fig.~\ref{fig:amplitude}); therefore for the
remainder of this paper we will describe our results using $A$.

\subsection{The role of convection}
\label{sec:res_conv} 
\begin{figure}[t]
  \begin{center}
    \includegraphics[angle=-90,width=8cm]{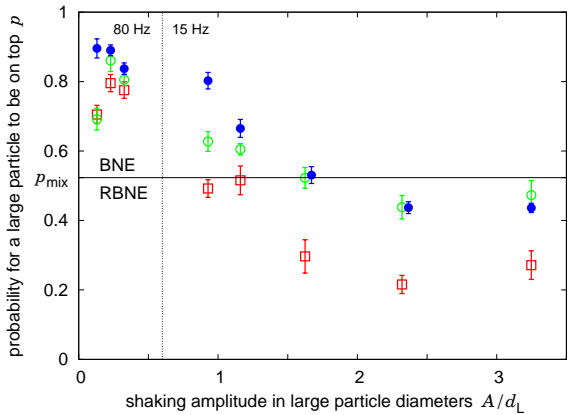}
    \caption{(Color online) At 15 Hz (to the right of the vertical dotted line):
    simulations with frictional sidewalls ({$\circ$}) agree with experiment ($\bullet$),
    while simulations
         with periodic boundary conditions ({\tiny$\square$})  differ. This shows the
         importance of side wall driven convection rolls at this frequency.
    At 80 Hz (to the left of the vertical dotted line):
    the simulations are independent of the boundary conditions.
    The experimental results are the same as in
        Fig.~\ref{fig:amplitude}.
        All simulation results are averaged over 10 runs starting from different initial conditions.
        }
    \label{fig:convection_md}
  \end{center}
\end{figure}

Convection driven by the sidewalls of the
container cannot be switched off in experiment. To estimate the
relative importance of this mechanism we therefore rely on the
molecular dynamics simulations described in Sec.~\ref{sec:MD},
where we can suppress convection by using periodic boundary
conditions for the lateral walls of the container
\cite{ciamarra:05}. Figure  \ref{fig:convection_md} shows again
good agreement between experiment and simulations with frictional
sidewalls for all but the smallest shaking amplitude at both 15 Hz
and 80 Hz. However, the role of convection is different at the two
frequencies.

Switching off convection in the simulations at 15 Hz results in a
significantly lower probability for a large particle to be on top.
For shaking amplitudes smaller than 1.2 $d_{\rm L}$, switching off
convection makes the system go from BNE to  a mixed state. At
higher amplitudes another RBNE mechanism eventually becomes
stronger than convection so that the overall effect even with
convection becomes RBNE.

Simulation results at 80 Hz with frictional sidewalls and periodic boundaries agree quite well,
showing that convection is not responsible for the strong BNE observed there.
This agrees with the experimental observations:
in high-speed movies of the top and bottom surface of the sample the
driving mechanism was identified as void filling.

This dependence of the segregation mechanism on $f$ agrees with the previous experimental observations:
Vanel {\it et al.}~\cite{vanel:97}
found the motion of an intruder to be governed by convection for frequencies below 40 Hz
and by void filling above. Other experiments observing convection were performed with well separated
taps \cite{knight:93,knight:96,moebius:01,moebius:04,moebius:05} or at 10 Hz \cite{cooke:96}.

\begin{figure}[t]
  \begin{center}
    \includegraphics[angle=-90,width=8cm]{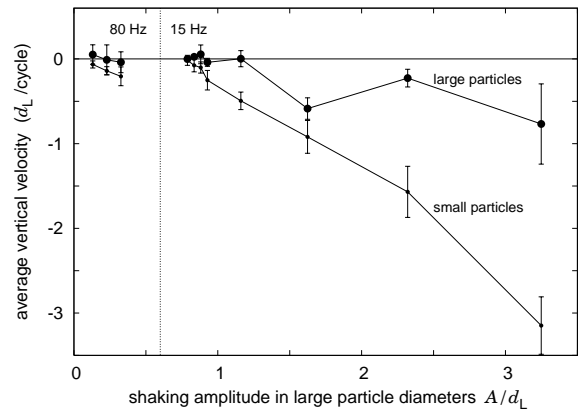}
    \caption{Convection becomes important as the shaking amplitude increases.
    These measurements, performed using high-speed images as in
    Fig.~\ref{fig:high_speed}, show the downward velocity of the particles at the sidewalls of the container.
        The sample is the same as in Fig.~\ref{fig:driving}.
        The velocities are averaged over all phases of 6 cycles (15 Hz) or 30 cycles (80 Hz).
        }
    \label{fig:conv_1}
  \end{center}
\end{figure}

Figure \ref{fig:conv_1} shows the experimentally measured average vertical velocities
in the layer adjacent to the sidewalls. For small  $A$ the observed velocities are
comparable in size to the measurement errors.
 However, for $f$ = 15 Hz and $A > d_{\rm L}$, the small particles start to become
dragged down significantly, which is in agreement with the findings
in Fig.~\ref{fig:convection_md}. For shaking amplitudes larger
than 1.5  $d_{\rm L}$,  large particles also get entrained in the
downward flow. At these  amplitudes the system shows a slight
RBNE, so we are unable to decide if convection still drives the
system towards BNE or if it has become a mixing mechanism.

\subsection{The granular temperatures of the two species}
\label{sec:md_gran_temp}

\begin{figure}[t]
  \begin{center}
    \includegraphics[angle=-90,width=8.5cm]{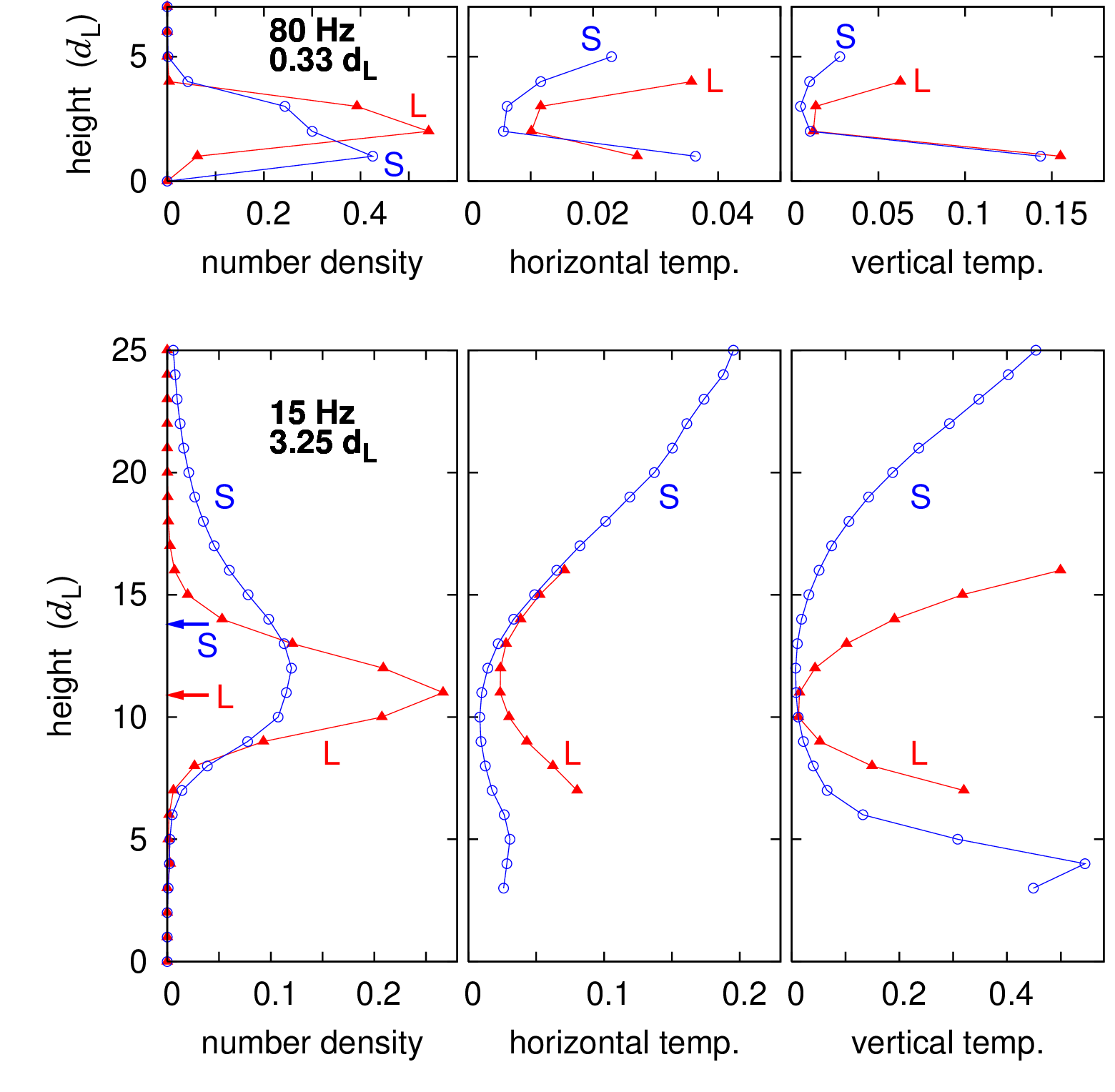}
    \caption{(Color online) Number density and granular temperature in vertical and horizontal directions
        in simulations with periodic boundaries.
    As predicted by thermal diffusion, the larger particles
        accumulate preferentially at the minimum of the granular temperature.
    The sample contains one mono layer (measured in units of $d_{\rm L}$) each of small ({$\circ$}) and
    large ({$\blacktriangle$}) particles
        with a diameter ratio of two.
    The RBNE observed in the experiment at 15 Hz is in accord with the 
	relative positions of the arrows, which show the height of the the centers of mass
    for the small and large particles. The granular temperatures were only computed for
    heights where there was in average at least one particle in a height intervall of $d_{\rm L}$
    of the given type at a time;
    their values are scaled by the energy needed to raise a large particle its own diameter.
        Zero height corresponds to the lowest position of the vibrating plate.
        Data are averaged at a single phase of the cycle for 1200  (80 Hz) or  434 (15 Hz) cycles.
    At 15 Hz this phase corresponds to the most expanded sample at the upper turning
    point; at 80 Hz no dependence on the phase was observed.
      }
    \label{fig:gran_temp}
  \end{center}
\end{figure}

Figure \ref{fig:gran_temp} shows the height dependence of the
number density and the granular temperature in  simulations with periodic boundaries.
Because gravity and shaking differentiates between vertical and horizontal motion, we
analyze the granular temperature in these two directions separately.
The upper frame characterizes the
fields  at $f$ =  80 Hz and $A$ = 0.33 $d_{\rm L}$; their behavior is independent of the phase of the driving.
The center of mass of the large particles is higher than that of the small particles,
corresponding to the BNE found after switching off the shaking.
While this BNE is generally attributed to the void-filling mechanism, it is interesting that the center
of mass of the large particle has also the same height as the minimum of granular temperature in both
horizontal and vertical directions. This hints that the realm of thermal diffusion and therefore
kinetic theory might extend to relatively dense samples.

At $f$ = 15 Hz and $A$ = 3.25 $d_{\rm L}$ the results do depend on
the phase of the shaking cycle. However, after the initial shock
wave induced by the plate hitting the sample has traveled upwards
\cite{bougie:02}, the remaining 5/6 of the shaking cycle is well
represented by the data in Fig.~\ref{fig:gran_temp}. Two features
are characteristic for this ``free flight'' phase: first the
center of mass of the large particles is always below that of the
small particles;
this agrees with the RBNE found in experiment  and simulation.
Second, the center of mass of the large particles is
at approximately the same height as the minimum of granular
temperature in both horizontal and vertical directions.

These two features point clearly to the thermal diffusion mechanism
suggested by Hsiau and Hunt \cite{hsiau:96}
to be  the cause for the observed RBNE.
Minima of granular temperature are a generic feature of vertically
vibrated granular samples \cite{soto:99,moon:04,feitosa:02,huan:04},
so thermal diffusion can be expected to be a relevant
segregation mechanism in many systems.

Characterizing the degree of non-equipartition is difficult due to the strong spatio-temporal dependencies of
the granular temperatures of both species. For a first estimate we have computed a weighted ratio:
\begin{equation}
\label{eq:temp_ratio}
\frac {T_{\rm L}}{T_{\rm S}} = \sum_z  n_{\rm L}(z)   \frac {T_{\rm L}(z)}{T_{\rm S}(z)},
\end{equation}
where $ n_{\rm L}(z)$ is the height dependent number density of large particles.
Averaged over the whole cycle, the simulations at $f$ = 15 Hz and $A$ = 3.25 $d_{\rm L}$
yield a temperature ratio of 5.4 $\pm$ 2.2 in vertical direction and 2.8 $\pm$ 0.6 in horizontal
direction. The larger standard deviation in vertical direction results from a threefold increase
of $T_{\rm L} / T_{\rm S}$ during the initial shock wave.
At $f$ = 80 Hz and $A$ = 0.33 $d_{\rm L}$
the temperature ratio is 2.2  in vertical direction and 1.7  in horizontal
direction.

Predictions of $T_{\rm L} / T_{\rm S}$ from kinetic theory assume
a dilute granular gas either driven by a stochastic thermostat
(Barrat and Trizac \cite{barrat:02}) or shaken at high frequencies
and high amplitudes (Brey {\it et al.}~\cite{brey:05}). Figure
\ref{fig:high_speed} demonstrates that a granular gas is most
closely realized by shaking with $f$ = 15 Hz and $A$ = 3.25
$d_{\rm L}$. With our diameter ratio, coefficient of restitution,
and number of particles, the polynomial equation given by Barrat
and Trizac predicts $T_{\rm L} /T_{\rm S} $ = 2.62, which is in
good agreement with our result  of 2.8 $\pm$ 0.6 for the
horizontal temperature ratio (which  depends less on the phase of
the shaking cycle). The cubic equation for the ratio of the mean
square velocities given by Brey {\it et al.}~\cite{brey:05}
results in $T_{\rm L} /T_{\rm S}$ = 1.28; however, their result
was derived for the case of a single large intruder, not equal
volumes of small and large particles. We will return to this
question in Sec.~\ref{sec:diameter_ratio}.


\subsection{Relevance of the different mechanisms}
\label{sec:results_mechanism}

The following overall picture emerges: at high frequencies and
small shaking amplitudes the sample shows a strong BNE due to {\bf
void filling}. At low frequencies and small shaking amplitudes,
sidewall driven {\bf convection} also leads to a BNE. However,
with increasing shaking amplitude, {\bf thermal diffusion} becomes
relevant: because the minimum of granular temperature is lower
than the center of mass of the whole sample, thermal diffusion results
in a RBNE that eventually becomes stronger than the BNE due to
convection.

By Eqs.~\ref{eq:epsilon_cond} and \ref{eq:epsilon_vf},
the control parameters for both {\bf condensation}
and void filling have the same numerical value of 8. Therefore, the theory of
Hong {\it et al.} predicts a strong BNE except for some intermediate
driving range where condensation takes place and is of equal strength as  void filling, which should lead to
a mixed state. Figure \ref{fig:high_speed} shows that a condensation of large particles is only conceivable
for shaking amplitudes smaller than 1.16 $d_{\rm L}$. Neither the experimental data in Fig.~\ref{fig:amplitude}
nor the simulations in  Fig.~\ref{fig:convection_md} show signs of condensation taking place at any
frequency and amplitude.

The effect of {\bf non-equipartition} and  {\bf static compressive
force} on the RBNE at high shaking amplitudes can be more
appropriately studied by their dependence on the diameter ratio,
as described in the next section.

\section{Results II: Influence of diameter ratio, volume ratio,  and total layer height}
\label{sec:results_2}
\subsection{Dependence on diameter ratio}
\label{sec:diameter_ratio}
\begin{figure*}[t]
  \begin{center}
    \includegraphics[angle=-90,width=16cm]{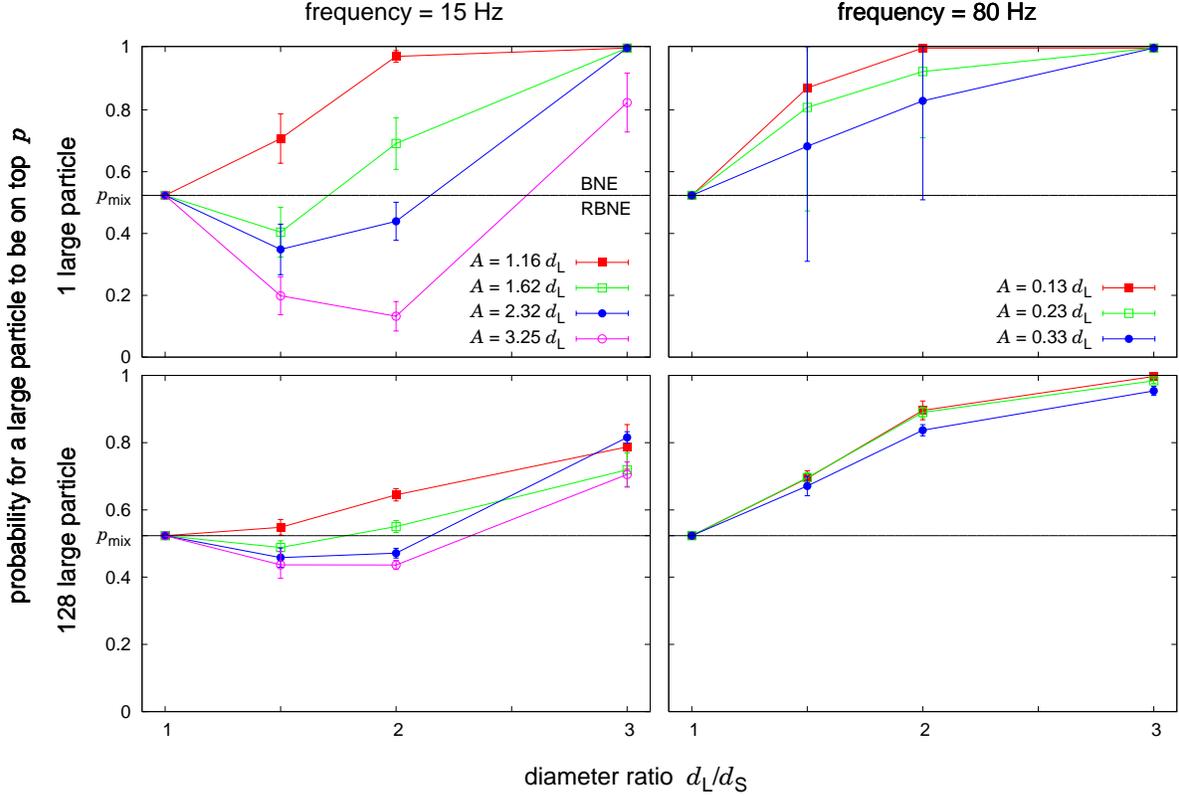}
    \caption{(Color online) Dependence of the observed segregation on the diameter ratio.
        The samples in the upper plot contain a single large particle
        in 2 layers (measured in $d_{\rm L}$) of small particles.
        The samples in the lower plot consist of equal volumes of each 1 monolayer of
        small and large particles.
        All data points  are averaged over 19 independent experiments.}
    \label{fig:dr}
  \end{center}
\end{figure*}

The dependence of the experimental segregation results on the diameter ratio is shown in Fig.~\ref{fig:dr}.
At $f$ = 80 Hz a change of the diameter ratio  from 1.5 to 3 always increases the BNE, which is in agreement
with our interpretation of void filling as the relevant mechanism.

At $f$ = 15 Hz the situation is more complex: for  $A$ = 1.16 $d_{\rm L}$,
the system exhibits BNE; its strength increases with
diameter ratio. At larger shaking amplitudes  we observe RBNE at
small and BNE at larger diameter ratios. This general behavior is
the same for a single large intruder and equal volumes of large
and small particles; however, the effects  are more pronounced for
the single intruder.

\begin{figure}[t]
  \begin{center}
    \includegraphics[angle=-90,width=8.5cm]{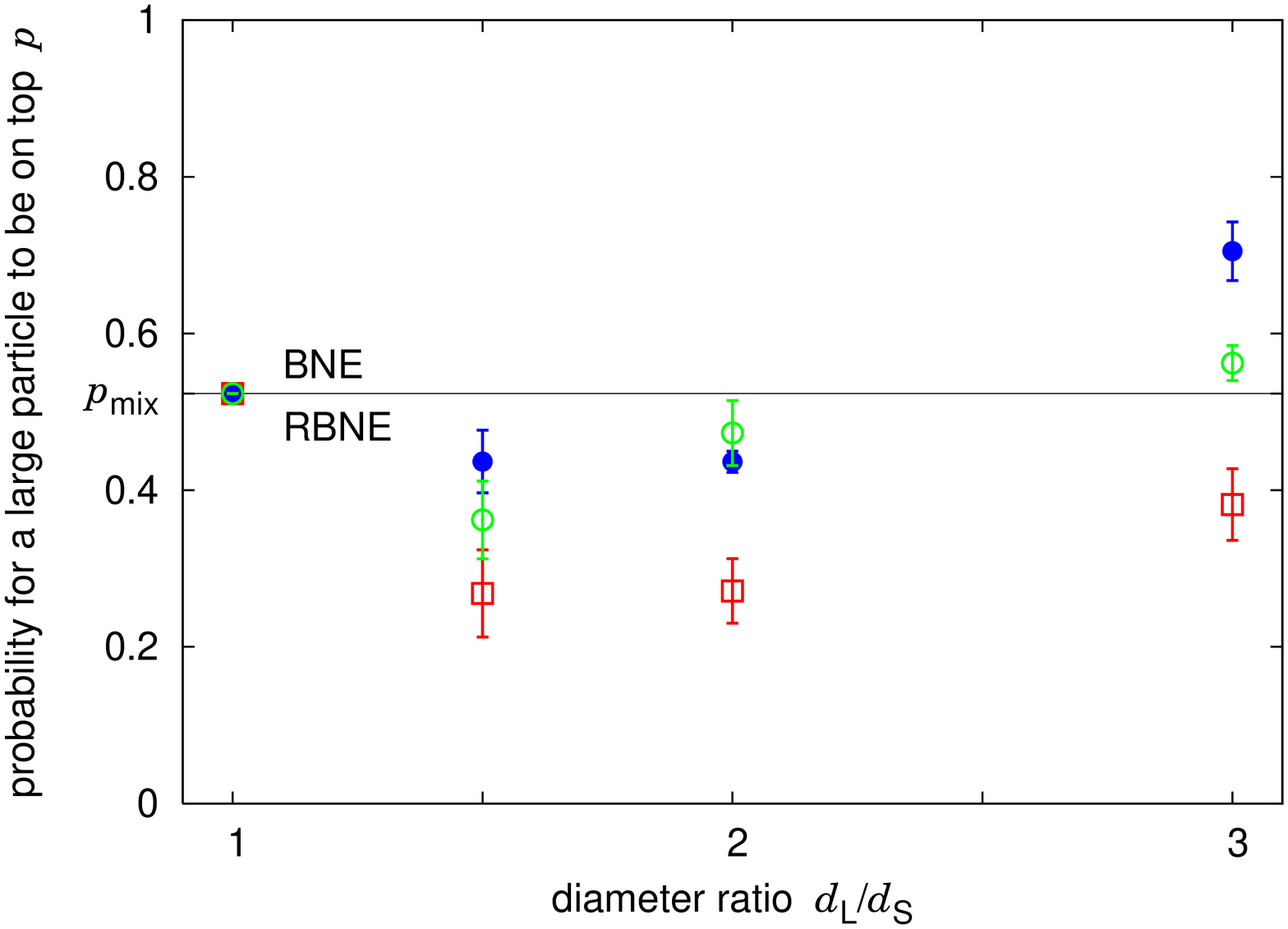}
    \caption{(Color online) Dependence of the segregation  on the diameter ratio
        with and without convection.
        The closed circles ($\bullet$) correspond to the experimental results for
    $f$ = 15 Hz, $A$ = 3.25 $d_{\rm L}$ and equal volumes of small and large particles(Fig.~\ref{fig:dr}).
    The open circles ({$\circ$}) are the results of the MD
        simulations using frictional sidewalls.
        The open squares ({\tiny$\square$}) represent the simulation results if
        periodic boundary conditions are used to suppress convection.
        All simulation results are averaged over 10 runs starting from different initial conditions.
        }
    \label{fig:dr_md}
  \end{center}
\end{figure}

We know from Fig.~\ref{fig:convection_md} and \ref{fig:gran_temp}
that the results at $f$ = 15 Hz involve at least two mechanisms:
convection and thermal diffusion. Since we want to use the dependence on the diameter ratio
to study the role of non-equipartition and static compressive force, we first try to isolate
the contribution of convection. In analogy to Fig.~\ref{fig:convection_md},
we do this  by comparing the experimental results with
simulations with rigid and periodic boundaries, as shown in Fig.~\ref{fig:dr_md}.
We consider only the case of the largest shaking amplitude because it can be expected to
resemble best the driving assumed in \cite{trujillo:03,brey:05}.

The agreement in Fig.~\ref{fig:dr_md} between the experimental results and the simulations with rigid boundaries
decreases for diameter ratios of 1.5 and 3. This is probably due to two reasons: First we kept the
collision parameters in the  simulation at the values obtained from the fit with data for a  diameter ratio of two;
however, the coefficient of restitution is known to change with the particle diameter \cite{brilliantov:04}.
Second, the large size discrepancy leads to increased probability of multi-particle collisions, thus the $ \rm T_c$
model of Luding and McNamara was implemented for the simulations with a diameter ratio of 3 \cite{luding:98}.
In this scheme, the restitution coefficient was set to 1 if a particle encountered a second collision
within a short time frame. Still only a subset of the simulations ran successfully.

\begin{table} [b]
\caption{Comparison of the ratio of granular temperatures $T_{\rm L} /T_{\rm S}$
between our simulations
and the kinetic theory based predictions by Barrat and Trizac \cite{barrat:02}.
The samples consist of 1 layer of each of small and large particles and are shaken at
$f$ = 15 Hz and  $A$ = 3.25 $d_{\rm L}$.  The simulations use periodic boundary conditions;
$T_{\rm L} /T_{\rm S}$ is calculated using Eq.~\ref{eq:temp_ratio} and the horizontal temperatures.}
\label{tab:temp_ratio}
\begin{tabular}{l||c|c|c}
$d_{\rm L}/ d_{\rm S}$ & 1.5  & 2 & 3 \\
\hline
\hline
Barrat and Trizac & 1.51 & 2.62 & 9.11 \\
\hline
MD simulation & 1.3 $\pm$ 0.1 & 2.9 $\pm$ 0.6 & 8.9 $\pm$ 3.8 \\
\end{tabular}
\end{table}

\begin{figure} [t]
  \begin{center}
    \includegraphics[angle=-90,width=8.5cm]{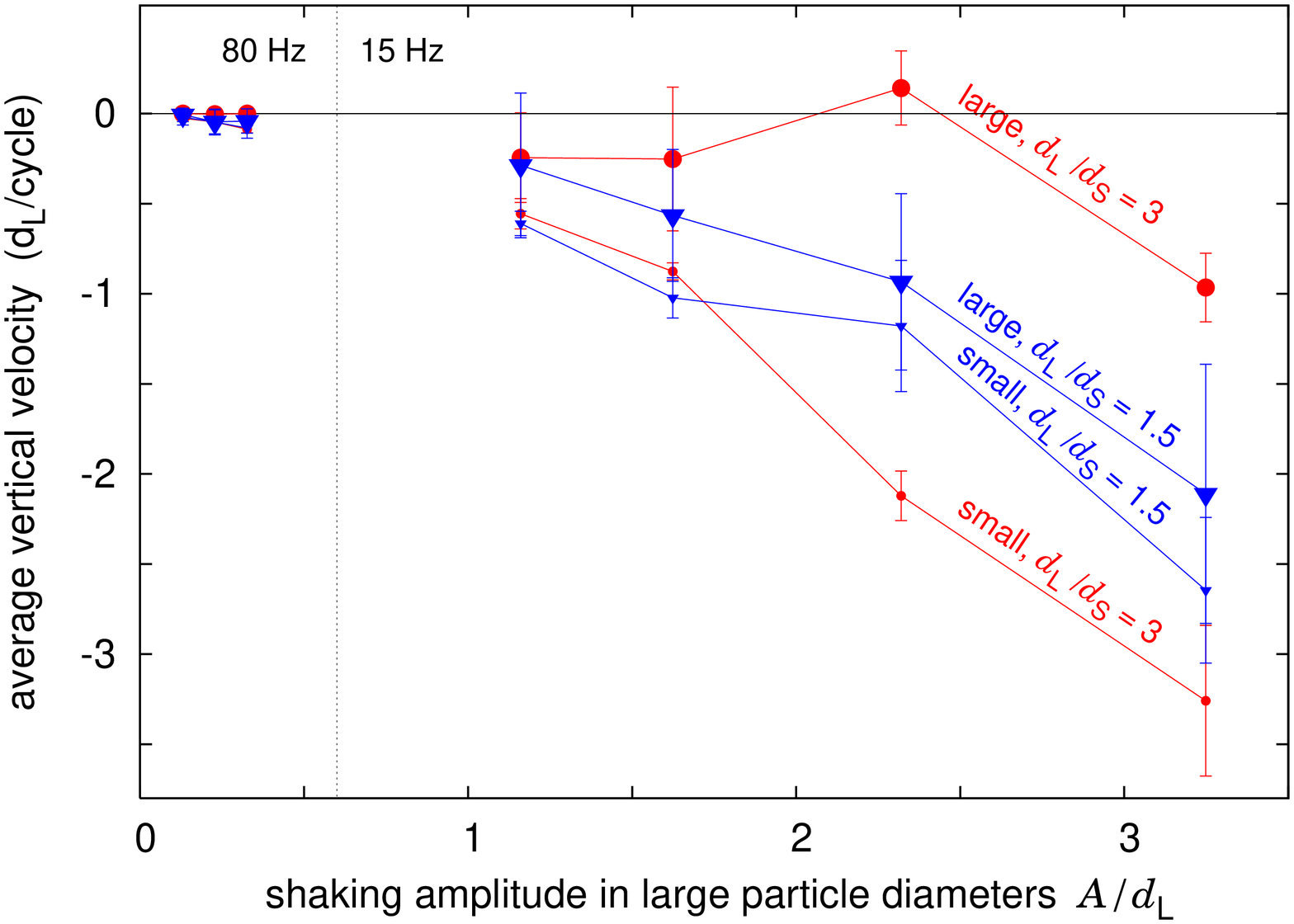}
    \caption{(Color online) Segregation due to sidewall driven convection increases with diameter ratio.
        The mean vertical velocity for small and large particles was
        measured in experiment at the sidewall of the container
        for diameter ratios of 1.5 ($\blacktriangledown$) and 3 ($\bullet$).
        Small and large symbols represent the respective particles.
        Both samples contain equal volumes of small and large particles.
        The velocities are averaged over all phases of 6 cycles (15 Hz) or 30 cycles (80 Hz.)
        }
    \label{fig:dr_conv}
  \end{center}
\end{figure}

Despite these caveats Fig.~\ref{fig:dr_md} shows that the BNE for a diameter ratio of 3 is due to an increase
in convection.
This effect can also be seen in the average particle velocity measured at the sidewall of the container in
Fig.~\ref{fig:dr_conv}. At small shaking amplitudes the convection is again negligible as in
Fig.~\ref{fig:conv_1}. For $A > d_{\rm L}$ the difference in the downward velocity for small and large
particles increases significantly with the diameter ratio.

Aside from the effect of convection, the main result in
Fig.~\ref{fig:dr_md} is that the slight RBNE in strongly shaken
mixtures is relatively independent of the diameter ratio. Yet
non-equipartition increases by a factor of six for the diameter
ratios studied, as shown in Table \ref{tab:temp_ratio}, which also
shows that $T_{\rm L} /T_{\rm S}$ is in good agreement with the
kinetic theory based predictions of Barrat and Trizac
\cite{barrat:02}.

Together this relative independence of $p$ on $d_{\rm L}/ d_{\rm S}$ and the strong dependence
of $T_{\rm L} /T_{\rm S}$ on $d_{\rm L}/ d_{\rm S}$
indicates that non-equipartition has only a weak influence on our results.
This is in contrast to the predictions of Trujillo \emph{et al.}~\cite{trujillo:03}, who
expect the static compressive force which leads to RBNE to be dominant at small diameter ratios.
However, non-equipartition grows faster with increasing diameter ratio than the static compressive force,
so the system should show a crossover to BNE.

For the reasons given above, we were unable to gather meaningful
statistics for simulations of a single intruder in a container with periodic boundaries.
Therefore, we cannot  comment on the predictions of Brey {\it et
al.}~\cite{brey:05}, who found for an intruder that
non-equipartition leads to a RBNE that increases in strength with
increasing diameter ratio.

\subsection{Dependence on the volume ratio $V_{\rm L} /V_{\rm S}$}

\begin{figure*} [t]
  \begin{center}
    \includegraphics[angle=-90,width=16cm]{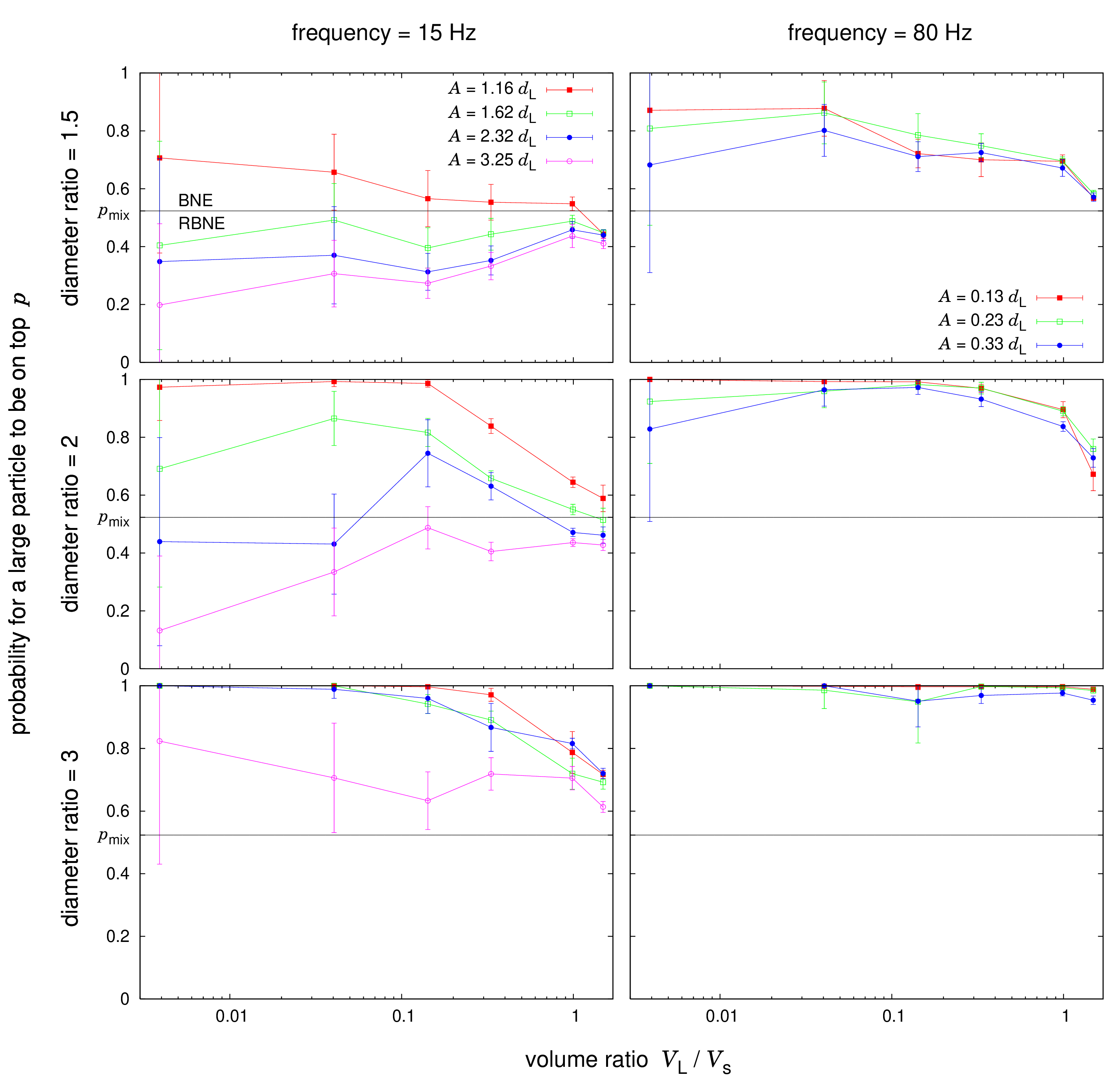}
    \caption{(Color online) Dependence of the observed segregation  on the volume ratio between small
        and large particles. All samples have a total volume of 2 layers (measured in $d_{\rm L}$).
        The left-most  data points in each frame correspond to a single large intruder.
        All results are averaged over at least 18 independent experiments.
        }
    \label{fig:2ml}
  \end{center}
\end{figure*}

\label{sec:height_ratio}
The dependence of the segregation results on the composition
of the sample measured as the ratio between the volumes of the large and small particle
 $V_{\rm L} / V_{\rm S}$ is shown in Fig.~\ref{fig:2ml}.
For each volume ratio the same behavior as in
Fig.~\ref{fig:amplitude} can be observed: the probability $p$ of a
large particle to be at  the top of the sample decreases with
increasing $A$. For volume ratios in the range from a single
intruder up to 20\% large particles, $p$ depends only weakly on
$V_{\rm L} / V_{\rm S}$.

At higher volume ratios, $p$ shows a trend from BNE towards a mixed state with increasing
 $V_{\rm L} / V_{\rm S}$. This effect is in agreement with the suggested segregation mechanism:
at $f$ = 15 Hz the size of the convective downstream layer increases and
 more large  particles can get entrained \cite{kudrolli:04}, while
 at $f$ = 80 Hz an increase of  $V_{\rm L} / V_{\rm S}$ results in larger voids in the sample
and therefore the void-filling  mechanism becomes less
effective.  A comparison of the two frequencies shows that
convection is more impaired by an increase in $V_{\rm L} / V_{\rm
S}$ than void filling.

An increasing RBNE with increasing diameter ratio
was not found for any of the combinations of volume ratios and
driving parameters in Fig.~\ref{fig:2ml}.  Therefore, a static
compressive force is not needed to explain our results.

Figure \ref{fig:2ml} provides another test of the theory of Hong
{\it et al.} \cite{hong:01}. According to Eq.~\ref{eq:HQL_1} the
condensation mechanism leading to RBNE will become stronger than
the void filling mechanism for volume ratios larger than 1. The
right-most points in Fig.~\ref{fig:2ml} have $V_{\rm L} / V_{\rm
S}$ = 1.5; therefore, we should see here at some intermediate
driving regime a clear drop of from a BNE to RBNE and back to BNE
for a further increase in $A$.

\begin{figure}[t]
  \begin{center}
    \includegraphics[,angle=-90,width=8.5cm]{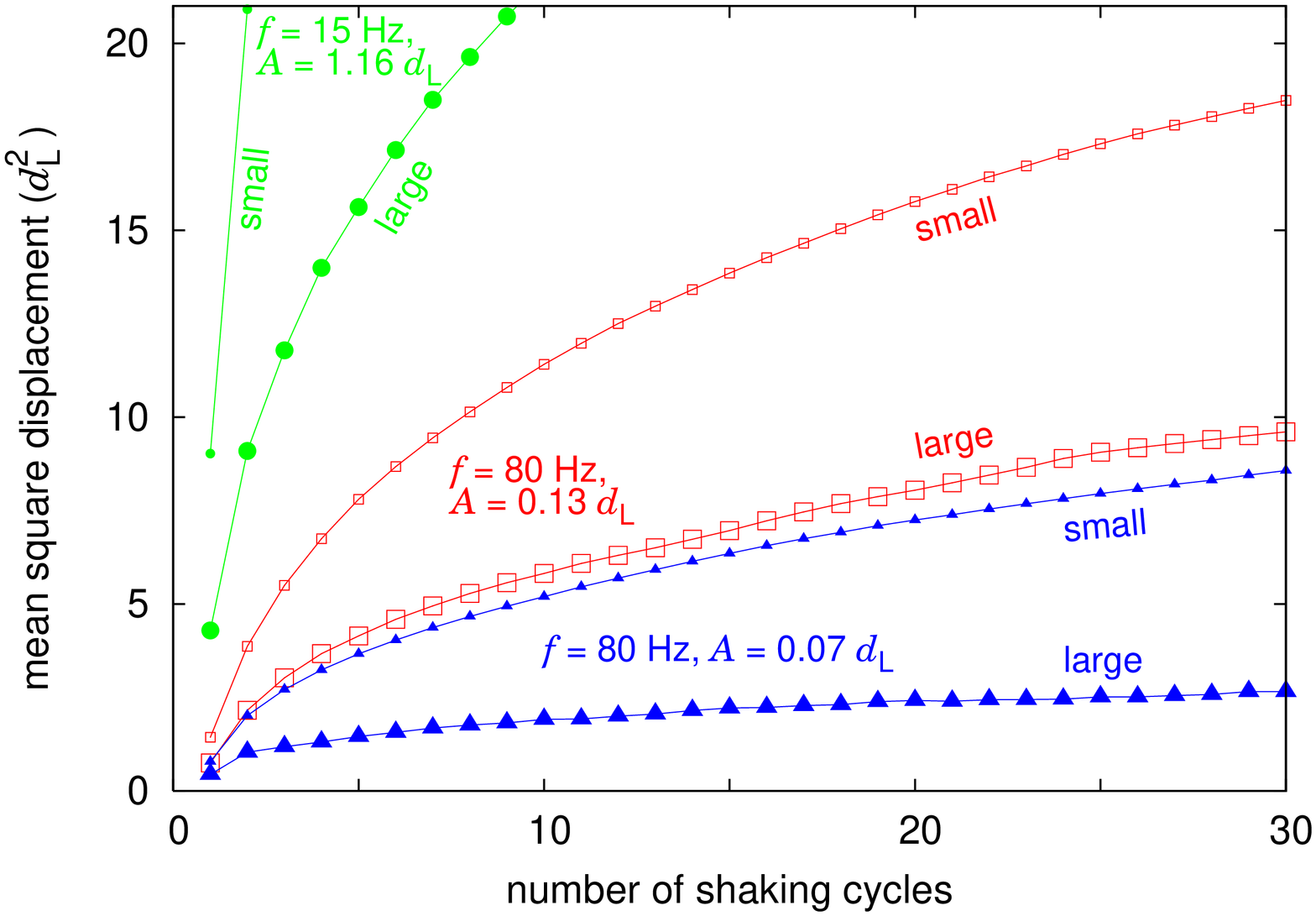}
    \caption{(Color online)
        Driving parameters adequate for the condensation mechanism were searched in
        simulations (with periodic boundary conditions) that yielded this plot of the mean
        square displacement. The diameter ratio is 2; the symbol size corresponds to the
particle type. The sample size is 2 layers with a volume ratio of
large to small spheres of 1.5.
    All data points are averaged over 200 time steps.
        }
    \label{fig:msd}
  \end{center}
\end{figure}

While no condensation was found at an intermediate
driving range, this leaves open the possibility of condensation at
even smaller driving amplitudes. Hong {\it et al.} consider a particle as
being fluidized if it has the ability to exchange its position
with its neighbors. Therefore, the mean square displacement of
particles in a condensed phase should show a plateau at a small
distance. Figure \ref{fig:msd} shows the development of the mean
square displacement of large and small particles in simulations
with periodic boundary conditions. Only at a frequency of 80 Hz
and an amplitude of 0.07 $d_{\rm L}$ (which pushes  the limits of
our simulation technique as discussed in
Sec.~\ref{sec:comparison}) do the large particles seem to be caged
by their neighbors, while the small particles are still mobile.
According to the condensation mechanism, this should result in a
RBNE. However, in experiment we find a clear BNE with $p$ = 0.67
(at a $V_{\rm L} / V_{\rm S}$ of 1.5). This steady state is
reached after approximately 20 minutes (corresponding to $10^5$
shaking cycles) and therefore unlikely to be found in simulations.
Similar experiments at $f$ = 15 Hz and shaking amplitudes  down to
0.61 $d_{\rm L}$ do not yield any signature of the condensation
mechanism.

\subsection{Dependence on the total layer height}
\label{sec:total_height}
\begin{figure}[t]
  \begin{center}
    \includegraphics[angle=-90,width=8.5cm]{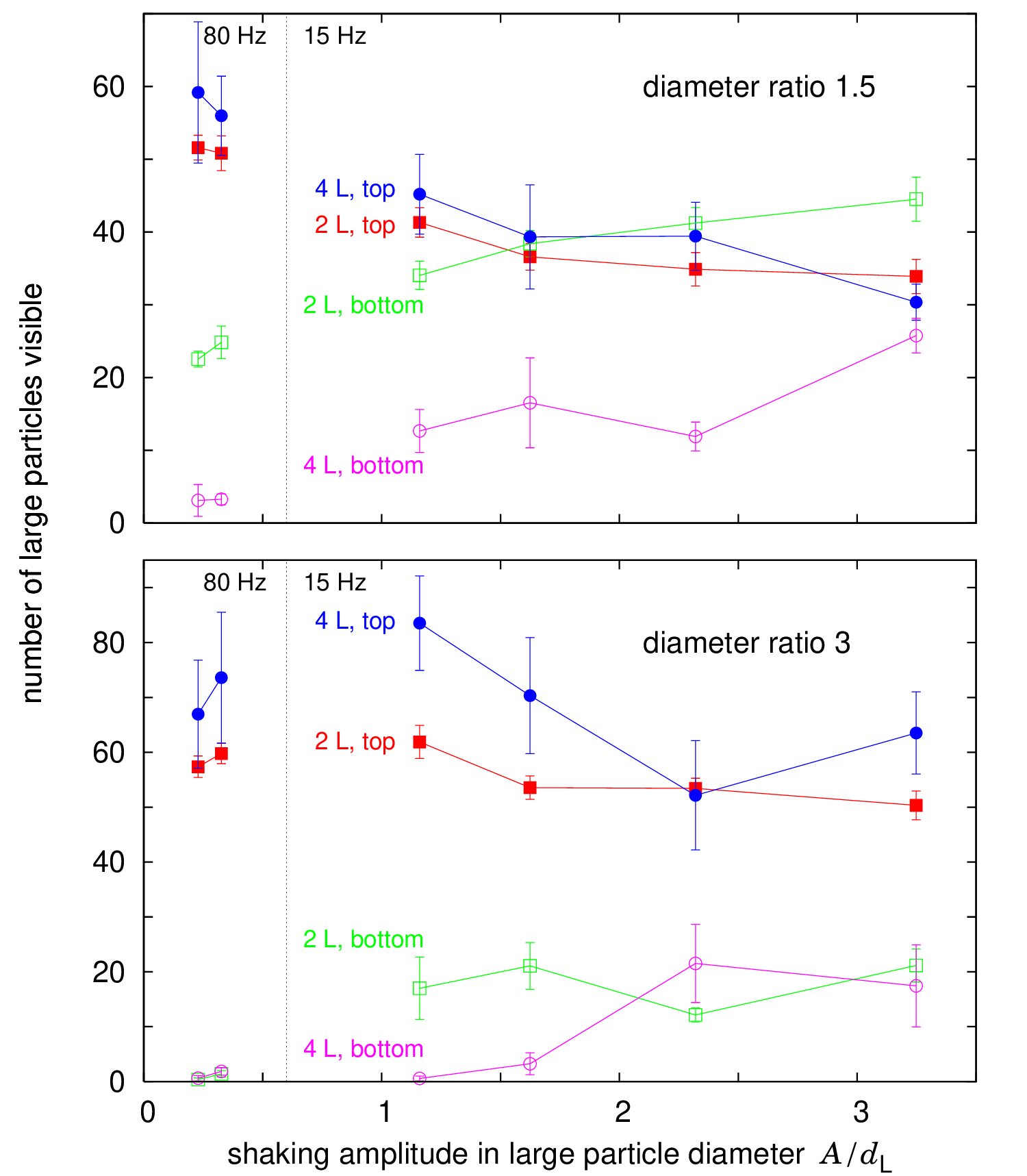}
    \caption{(Color online)
        Dependence of the observed segregation results on the total sample height,
        either 2 or 4 layers (measured in $d_{\rm L}$).
        All samples contain equal volumes of small and large particles.
        All data points are averaged over at least 12 independent
        experiments.
        }
    \label{fig:4ml}
  \end{center}
\end{figure}

The dependence of the segregation on the layer depth $h_{\rm total}$ is shown in
Fig.~\ref{fig:4ml}.
For a diameter ratio of 1.5 the number of large particles at the top  is mostly independent
of  $h_{\rm total}$ while the number at the bottom drops significantly with increasing $h_{\rm total}$.
At a diameter ratio of 3 there is a slight increase of the number at the top while the number at the bottom
stays relatively constant at low levels.

The absence of the RBNE in deeper layers was also noted by Breu
{\it et al.}~\cite{breu:03}. It is compatible with thermal
diffusion as the responsible mechanism for the RBNE in shallow
layers, assuming that the relative height of the temperature
minimum stays constant when $h_{\rm total}$ is increased.  Then in
deeper samples the accumulation of large particles would no longer
be visible at the bottom. The condensation mechanism
does not explain the  vanishing of the RBNE in deeper layers as
the condensed fraction of large particles is assumed to accumulate
at the bottom independent of $h_{\rm total}$. At a diameter ratio
of 3,  void filling and convection are much stronger contributions
to the final result; therefore, the relative influence of $h_{\rm
total}$ is smaller.

Note that the transition from two to four layers increases the relaxation times.
Though we increased the initial shaking time from 1 to 3 minutes, experiments with
$f$ = 80 Hz, $A$ = 0.13 $d_{\rm L}$ still had not reached their steady state and were therefore
not included.

\section{Conclusions}
\label{sec:conclusions} We have studied the segregation of
vertically shaken binary mixtures of brass spheres in experiment
and event-driven simulation.  By using the same material for small
and large particles and evacuating the sample container, we
eliminated Archimedean buoyancy, friction, and pressure effects
from the list of potential segregation mechanisms. We have found
that three mechanisms are sufficient to explain our results:

(1) At a shaking frequency of 80 Hz all samples exhibited  a
strong {\it Brazil nut effect} due to the geometrical mechanism
called {\bf void filling}. The strength of the segregation
increases with diameter ratio and decreases weakly with the
shaking amplitude. Increasing the number of large particles from a
single intruder up to a mixture of equal volumes of large and
small particles results only in a slight decrease of the strength
of the effect.

(2) At a shaking frequency of 15 Hz and shaking amplitudes up to
1.5 large particle diameters, a strong {\it Brazil nut effect} is
induced by {\bf side wall driven convection}. The strength of the
effect increases strongly with diameter ratio and decreases with
the relative amount of large particles.

(3) If the shaking amplitude is increased above 1.5 large particle
diameters, the sample starts to resemble a granular gas and the
{\bf thermal diffusion} mechanism becomes significant.
Thermal diffusion describes the tendency of large particles to
accumulate in the minimum of the granular temperature profile. In
a shallow layer of height two large particle diameters and for
diameter ratios smaller than three, thermal diffusion eventually
becomes stronger than convection, which leads to the {\it reverse
Brazil nut effect}. Thermal diffusion seems to be only weakly
dependent on the diameter ratio. If the total layer height is
increased to four large particle diameters, the temperature
minimum is within the sample and no increased number of large
particles at the bottom is observed.

While we probed  the parameter regimes where the {\bf static
compressive force} and the {\bf condensation} mechanisms were
expected to be relevant, we did not find any signature of these
two segregation mechanisms. The apparent agreement between the
condensation-based model in \cite{hong:01} and the experimental
phase diagram separating BNE and RBNE in \cite{breu:03} can be
explained by the fact that they changed both frequency and
amplitude of the driving deliberately. Then both BNE and RBNE
can occur in the same sample, as shown above and also in \cite{breu:03}.

We also find {\bf non-equipartition} to have no discernible
influence on our results. However, this might be due to either the
relatively high coefficient of restitution (0.7) or to the fact
that we are not shaking vigorously enough for those effects to be
observed.

The theories presented in \citep{hong:01,trujillo:03} lack two of the important mechanisms
(convection, thermal diffusion) and at the same time they include
mechanisms which are not relevant (static compressive force, non-equipartition, condensation);
hence these theories can not explain our experimental and
numerical results.

In the limit of strong shaking, we believe that kinetic theory-based
approaches have the potential to explain segregation,
though the inclusion of convection might prove daunting.
A good starting point for a microscopic
theory of void filling  might be the ``random fluctuating sieve``
theory of inclined chute flow \cite{savage:88}, as pointed out in
\cite{kudrolli:04}.

{\it Note added in proof.} The assumption in VI C that the
relative height of the temperature minimum is independent
of the total number of spheres is supported by the analysis in
Ref \cite{serero:06}.

Acknowledgments: We thank  Christine Hrenya, Javier Brey,
Massimo Pica Ciamarra, and Vicente Garz\'o  for helpful discussions,
the organizers of the  KITP program on Granular Physics 2005, and
Sibylle N\"agle for her help with Povray. This work was supported by the
Robert A Welch Foundation and by the
Engineering Research Program of the Office of Basic Energy Sciences of
the U.S. Department of Energy (Grant No. DE-FG03-93ER14312).



\begin{thebibliography}{63}
\bibitem{endnote}
Because the top surface is not flat and can therefore accommodate more particles, the
probability $p_{\rm mix}$  of a perfectly mixed state  corresponds to 0.523,
as determined from images of samples only consisting of big particles.
\end{thebibliography}

\end{document}